\documentclass[12pt,preprint]{aastex}

\begin{document}
 
\title{Differential Astrometry of Sub-arcsecond Scale Binaries
at the Palomar Testbed Interferometer} 
\author{B. F. Lane, M. W. Muterspaugh} 

\affil{Center for Space Research \\ MIT Department of Physics \\ 70 Vassar Street, Cambridge, MA 02139}
\email{blane@mit.edu}
\email{matthew1@mit.edu}

\begin{abstract} 

We have used the Palomar Testbed Interferometer to perform very high
precision differential astrometry on the 0.25 arcsecond separation
binary star HD 171779. In 70 minutes of observation we achieve a
measurement uncertainty of $\approx 9$ micro-arcseconds in one axis,
consistent with theoretical expectations. Night-to-night repeatability
over four nights is at the level of 16 micro-arcseconds. This method
of very-narrow-angle astrometry may be extremely useful for searching
for planets with masses as small as 0.5 $M_{Jup}$ around a previously
neglected class of stars -- so-called ``speckle binaries.''  It will
also provide measurements of stellar parameters such as masses and
distances, useful for constraining stellar models at the $10^{-3}$
level.

\end{abstract}


\section{Introduction}

Long-baseline optical interferometry promises high precision
astrometry using modest ground-based instruments. In particular the
Mark III Stellar Interferometer \citep{shao88} and Navy Prototype
Optical Interferometer (NPOI, Armstrong et al. 1998\nocite{arm98})
have achieved global astrometric precision at the ~10 mas (1 mas =
$10^{-3}$ arcseconds) level \citep{hum94}, while the Palomar Testbed
Interferometer (PTI, Colavita et al. 1999\nocite{col99} ) has
demonstrated an astrometric precision of 100 $\mu$as ($1 \mu$as =
$10^{-6}$ arcseconds) between moderately close (30 arcsecond) pairs of
bright stars \citep{shao92,col94,l00}.  While interferometric and
astrometric methods have proven very useful in studying binary stars,
and have long been argued to be well-suited to studying extra-solar
planets \citep{col94b,eis01}, to date results using these techniques
have been limited \citep{ben02b}.
  
There are several reasons why it is desirable to develop viable
astrometric planet-detection methods. Most importantly, the parameter
space explored by astrometry is complementary to that of radial
velocity (astrometry is more sensitive to larger separations). Second,
unlike current radial velocity detections, astrometric techniques can
be used to determine the orbital inclination of a planet. Finally,
astrometry is particularly well-suited to studying binary stellar
systems; such systems challenge other planet-finding techniques.  For
example, radial velocimetry can suffer from systematic velocity errors
caused by spectral contamination from the light of the second star
\citep{Vog:00::}. Similar problems are faced by coronographic
techniques, where the light from the second star is not usually
blocked by the occulting mask.

In this paper we describe recent efforts to obtain very high precision
narrow-angle astrometry using PTI to observe binary stars with
separations less than one arcsecond, i.e. systems that are typically
observed using speckle interferometry \citep{sah02} or adaptive
optics. Such small separations allow us to achieve astrometric
precision on the order of 10 $\mu$as, which for a typical
binary system in our target sample (binary separation of 20 AU),
should allow us to detect planets with masses down to 0.5 Jupiter
masses in orbits in the 2 AU range. This approach has been suggested
\citep{tcp96} and tried \citep{dyck95,chara03} before, though with
limited precision. However, this work is unique in that it makes use
of a phase-tracking interferometer; the use of phase-referencing
\citep{lc03} removes much of the effect of atmospheric turbulence,
improving the astrometric precision by a factor of order 100.

The Palomar Testbed Interferometer (PTI) is located on Palomar
Mountain near San Diego, CA \citep{col99}. It was developed by
the Jet Propulsion Laboratory, California Institute of Technology for
NASA, as a testbed for interferometric techniques applicable to the
Keck Interferometer and other missions such as the Space
Interferometry Mission, SIM.  It operates in the J ($1.2 \mu{\rm
m}$),H ($1.6 \mu{\rm m}$) and K ($2.2 \mu{\rm m}$) bands, and combines
starlight from two out of three available 40-cm apertures. The
apertures form a triangle with 86 and 110 meter baselines.

The paper is organized as follows: in Section 2 we describe the
experiment and derive expected performance levels. In Section 3 we
describe initial observations as well as the extensive data analysis
processing required to achieve the desired astrometric precision.
In Section 4 we discuss our preliminary results, and in Section 5 we
discuss the prospects of a larger search.

\section{Interferometric Astrometry}

In an optical interferometer light is collected at two or more
apertures and brought to a central location where the beams are
combined and a fringe pattern produced.  For a broadband source of
wavelength $\lambda$ the fringe pattern is limited in extent and
appears only when the optical paths through the arms of the
interferometer are equalized to within a coherence length ($\Lambda =
\lambda^2/\Delta\lambda$). For a two-aperture interferometer,
neglecting dispersion, the intensity measured at one of the combined
beams is given by
\begin{equation}
I(\delta) = I_0 \left ( 1 + V \frac{\sin\left(\pi \delta/ \Lambda\right)}
{\pi \delta/ \Lambda} \sin \left(2\pi \delta/\lambda \right ) \right )
\end{equation}
\noindent where $V$ is the fringe contrast or ``visibility'', which
can be related to the morphology of the source, and $\Delta\lambda$ is
the optical bandwidth of the interferometer assuming a flat optical
bandpass (for PTI $\Delta\lambda = 0.4 \mu$m). The differential
(between arms of the interferometer) optical path $\delta$ is found
from geometric considerations to be
\begin{equation}
\delta = \vec{B} \cdot \vec{s} - d
\end{equation}
\noindent $\vec{B}$ is the baseline -- the vector connecting the two
apertures. $\vec{s}$ is the unit vector in the source direction, and
$d$ is any differential optical path introduced by the instrument. For
a 100-m baseline interferometer an astrometric precision of 10 $\mu$as
corresponds to knowing $d$ to 5 nm, a difficult but not impossible
proposition.
 
The dominant source of measurement error is atmospheric turbulence
above the interferometer, which adds varying amounts of optical path
and hence makes the fringes appear to move about rapidly
\citep{rod81}. This atmospheric turbulence, which changes over
distances of tens of centimeters and millisecond time-scales, forces
one to use very short exposures to maintain fringe contrast, and hence
limits the sensitivity of the instrument. It also severely limits the
astrometric accuracy of a simple interferometer, at least over large
sky angles. However, this atmospheric turbulence is correlated over
small angles, and hence it is still possible to obtain high precision
``narrow-angle'' astrometry.
 
\subsection{Narrow-Angle Astrometry}
 
Dual-star interferometric narrow-angle astrometry \citep{shao92,col94}
promises astrometric performance at the 10-100 $\mu$as level for pairs
of stars separated by 10-60 arcseconds, and has been demonstrated at
PTI. Achieving such performance requires simultaneous measurements of
the fringe positions of both stars via the use of a ``dual-star''
optical beam-train.
 
For more closely spaced stars, it is possible to operate in a simpler
mode. We have recently used PTI to observe pairs of stars separated by
no more than one arcsecond. In this mode, the small separation of the
binary results in both binary components being in the field of view of
a single interferometric beam combiner. The fringe positions are
measured by modulating the instrumental delay with an amplitude large
enough to record both fringe packets. This eliminates the need for a
complex internal metrology system to measure the entire optical path
of the interferometer, and dramatically reduces the effect of
systematic error sources such as uncertainty in the baseline vector
(error sources which scale with the binary separation).

However, since the fringe position measurement of the two stars is no
longer truly simultaneous it is possible for the atmosphere to
introduce pathlength changes (and hence positional error) in the time
between measurements of the separate fringes. To reduce this effect we
split off a fraction of the incoming starlight and direct it to a
second beam-combiner. This beam-combiner is used in a
``fringe-tracking'' mode \citep{ss80,col99} where it rapidly (10 ms)
measures the phase of one of the starlight fringes, and adjusts the
internal delay to keep that phase constant.  This technique -- known
as phase referencing -- has the effect of stabilizing the fringe
measured by the astrometric beam-combiner. In addition, the residual
phase error measured by the fringe tracker is a high time-resolution
trace of the phase error introduced by the atmosphere and not fully
corrected by the fringe tracker; it can be applied to the measured
fringe position in post-processing.

\subsection{Expected Performance}

In making an astrometric measurement we modulate the optical delay
applied internally in a triangle-wave pattern around the stabilized
fringe position, while measuring the intensity of the combined
starlight beams. Typically we obtain one such ``scan'' every 1-3
seconds, consisting of up to 3000 intensity samples. The range of the
delay sweep is set to include both fringe packets; typically this
requires a scan amplitude on the order of $100~\mu$m.  We then fit a
double fringe packet based on Eqn. 1 to the data, and solve for the
differential optical path between fringe packets.

\begin{figure}[]
\begin{center}
\plotone{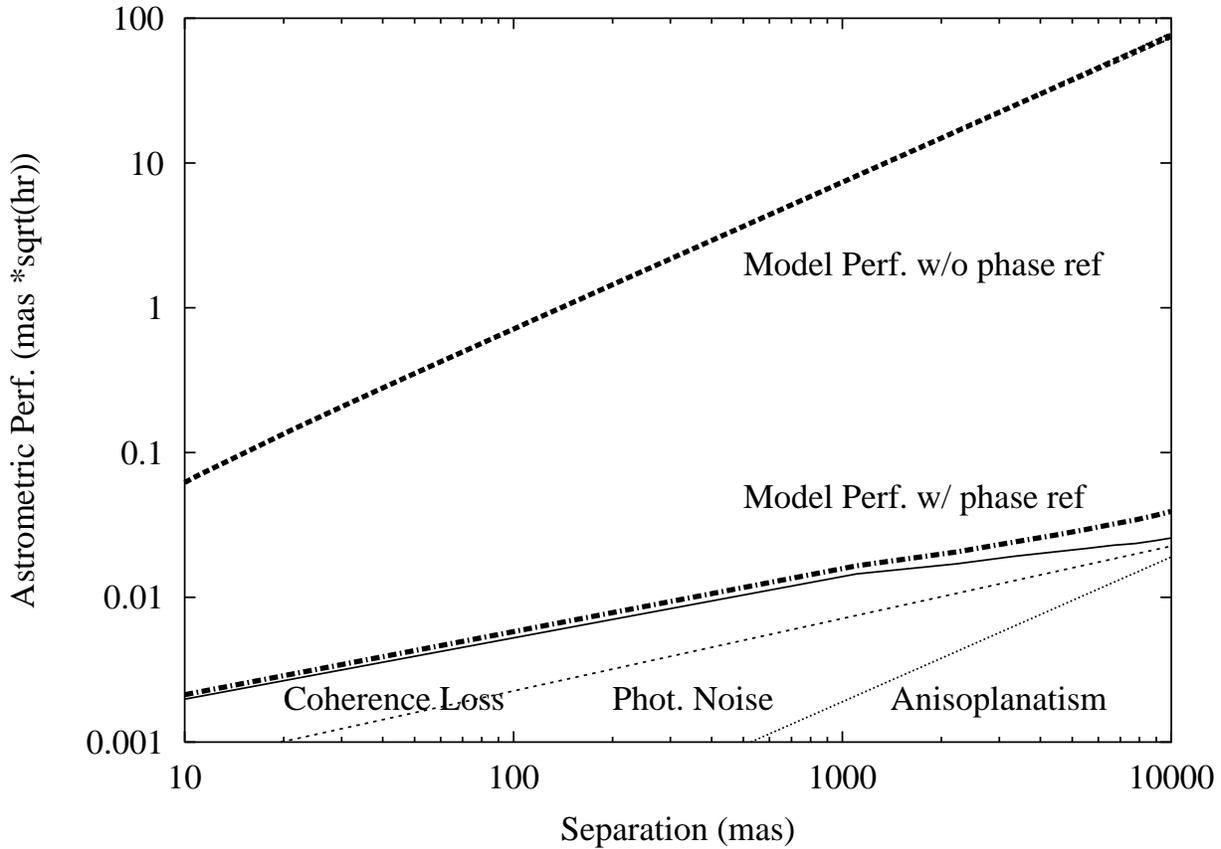}
\end{center}
\caption{\label{fig:expected} The expected narrow-angle astrometric
performance in mas for the phase-referenced fringe-scanning
approach, for a fixed delay sweep rate, and an interferometric
baseline of 110 m. The three error sources are described in Section 2.2.
Also shown is the magnitude of the temporal loss of coherence effect
in the absence of phase referencing, illustrating why stabilizing the
fringe via phase referencing is necessary.  }
\end{figure}

In calculating the expected astrometric performance we take into
account three major sources of error: errors caused by fringe motion
during the sweep between fringes (loss of coherence with time), errors
caused by differential atmospheric turbulence (loss of coherence with
sky angle, i.e. anisoplanatism), and measurement noise in the fringe
position. We quantify each in turn below, and the expected measurement
precision is the root-sum-squared of the terms (Figure
\ref{fig:expected}).

\subsubsection{Loss of Temporal Coherence}

\begin{figure}[]
\begin{center}
\plotone{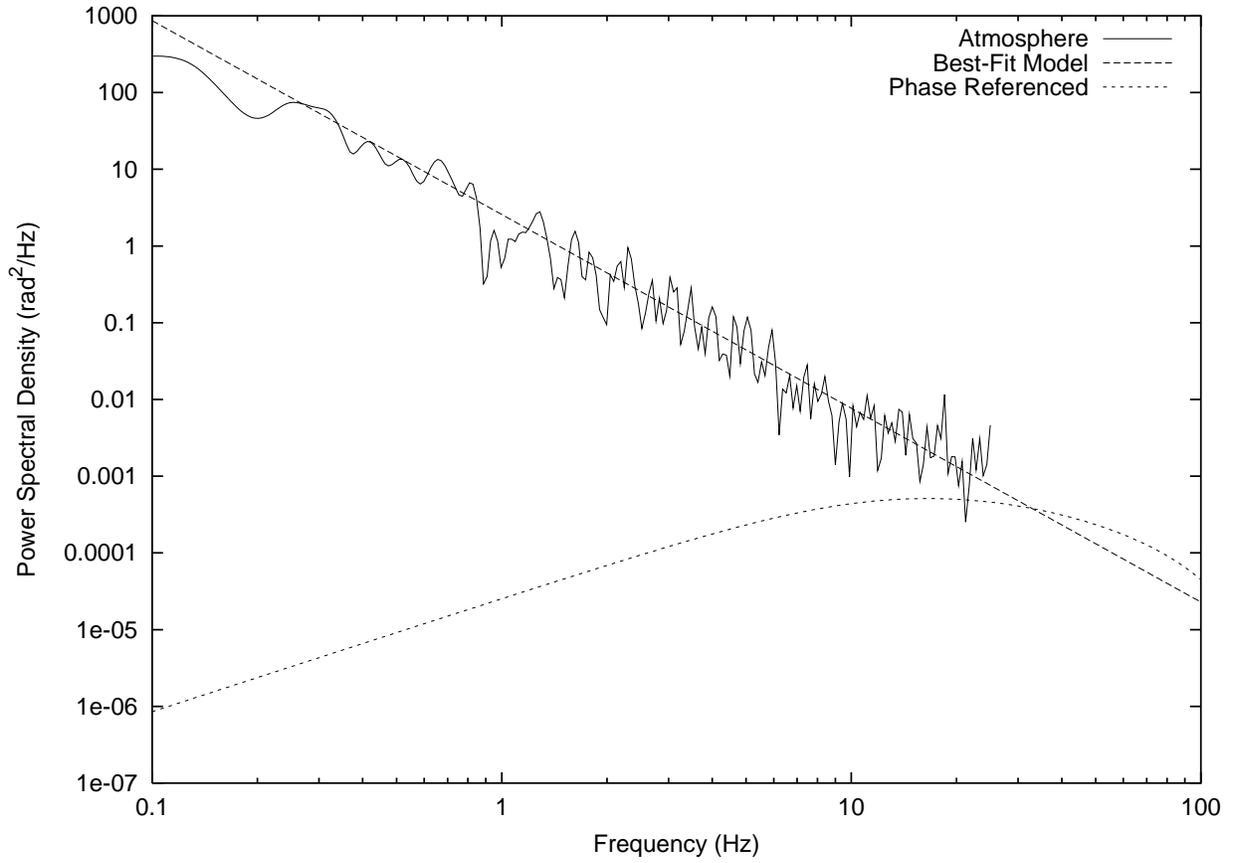}
\end{center}
\caption{\label{fig:pspec} Power spectral density of the fringe phase as measured by
PTI \citep{lc03}. The phase PSD is best fit by a power law $ A(f) \propto f^{-2.5}
$.  Also shown is the effective PSD of the phase noise after
phase referencing has stabilized the fringe.}
\end{figure}

The power spectral density of the fringe phase of a source observed
through the atmosphere has a power-law dependence on frequency (Figure \ref{fig:pspec});
at high frequencies typically
\begin{equation}
A(f) \propto f^{-\alpha}
\end{equation}
\noindent where $\alpha$ is usually in the range 2.5--2.7. The effect
of phase-referencing is to high-pass filter this atmospheric phase
noise. In our case, the servo is an integrating servo with finite 
processing delays and integration times, with the residual phase error 
``fed forward'' to the second beam combiner \citep{lc03}. The response of 
this system to an input atmospheric noise can be written 
in terms of frequency (see Appendix A in Lane 2003) as
\begin{eqnarray}
H(f)  & = &  \frac{1 - 2 {\rm sinc}(\pi f T_s)\cos(2 \pi f T_{d}) + {\rm sinc}^2(\pi f T_s) }{1 - 2 \frac{f_c}{f} {\rm sinc}(\pi f T_{s}) \sin(2 \pi f T_{d}) + \left ( \frac{f_c}{f} \right )^2 {\rm sinc}^2(\pi f T_{s})}
\end{eqnarray} 
\noindent where  $ {\rm sinc}(x) = \sin(x)/x$, $f_c$ is the closed-loop bandwidth
of the fringe-tracker servo (for this experiment $f_c = 10$ Hz), $T_s$ is the integration time
of the phase sample (6.75 ms), and $T_d$ is the delay between measurement
and correction (done in post-processing, effectively 5 ms). The 
phase noise superimposed on the double fringe measured by the 
astrometric beam combiner has a spectrum given by $A(f)H(f)$.

The sampling of the double fringe packet takes a finite amount of
time, first sampling one fringe, then the other. In the time domain
the sampling function can be represented as a ``top-hat'' function
convolved with a pair of delta functions (one positive, one negative).
The width of the top-hat is equal to the time taken to sweep through a
single fringe, while the separation between the delta functions is
equal to the time to sweep between fringes. In the frequency domain
this sampling function becomes
\begin{equation}
S(f) = \sin^2(2 \pi f \tau_p){\rm sinc}^2(\pi f \tau_\ast)
\end{equation}
where $\tau_p$ is the time taken to move the delay between stars
,$\Delta d/v_s$, and $\tau_{\ast}$ is the time to sweep through a
single stellar fringe, $\Lambda/v_s$. $v_s$ is the delay sweep
rate.

The resulting  error in the astrometric measurement, given in radians by $\sigma_{tc}$, can be found from
\begin{equation}
  \sigma_{tc}^2 = \left (\frac{\lambda}{2\pi {\rm B}} \right)^2 \frac{1}{N}\int_0^\infty A(f)H(f)S(f) df
\end{equation}
\noindent where ${N}$ is the number of measurements.  It is worth
noting that if phase-referencing is not used to stabilize the fringe,
i.e. $H(f) = 1$, the atmospheric noise contribution increases by a
factor of $\approx 10^2$--$10^3$.

\subsubsection{Anisoplanatism}

The performance of a simultaneous narrow-angle astrometric measurement
has been thoroughly analyzed in \cite{shao92}.  Here we
restate the primary result for the case of typical seeing at a site
such as Palomar Mountain, where the astrometric error in arcseconds 
due to anisoplanatism ($\sigma_{a}$) is
given by
\begin{equation}
\sigma_{a} = 540 {\rm B}^{-2/3} \theta t^{1/2} 
\end{equation}
\noindent where B is the baseline, $\theta$ is the angular separation of the
stars, and $t$ the integration time in seconds. This assumes a
standard \citep{l80} atmospheric turbulence profile; it is likely that
particularly good sites will have somewhat (factor of two) better
performance.

\subsubsection{Photon Noise}

The astrometric error due to photon-noise ($\sigma_{p}$) is given in radians as  
\begin{equation}
\sigma_{p} = \frac{\lambda}{2\pi {\rm B}} \frac{1}{\sqrt{N}} \frac{1}{{\rm SNR}}
\end{equation}
\noindent where N is the number of fringe scans, and SNR is the signal-to-noise 
ratio of an individual fringe.

\section{Observations \& Data Processing}

We used the Palomar Testbed Interferometer to observe the binary star
HD 171779 (HR 6983, K0III+G9III, $m_K = 2.78$, $m_V = 5.37$) on the
nights of 11-14 August 2003. Conditions were calm with good seeing
although with high-altitude clouds limiting observations, particularly
after the first night. Based on a visual orbit obtained from speckle
interferometry \citep{hart01}, the predicted separation was 0.248
arcsec at a position-angle of 120.9 degrees. The orbit quality was
listed as 3 (on a scale of 1--5), indicating a ``reliable'' orbit,
though no uncertainties are given. The published $\Delta V = 0.21$
\citep{wor01}, and the orbital period is 191.49 years.

We used the longest available baseline (110 m). The observations were
done in the K band ($2.2 \mu$m), with $\approx$70\% of the K-band
light being used for fringe tracking, and the remaining light going to
the astrometric measurement. The fringe tracker operated in the
standard configuration \citep{col99}, with a sample time of 10 ms and
a closed-loop servo bandwidth of $\approx 10$ Hz. The delay modulation
used for the astrometric measurement typically had an amplitude of 150
$\mu$m and period of 3 seconds. The modulation was done using a
PZT-actuated mirror, the position of which was measured using a laser
metrology system.

\begin{figure}[]
\begin{center}
\plotone{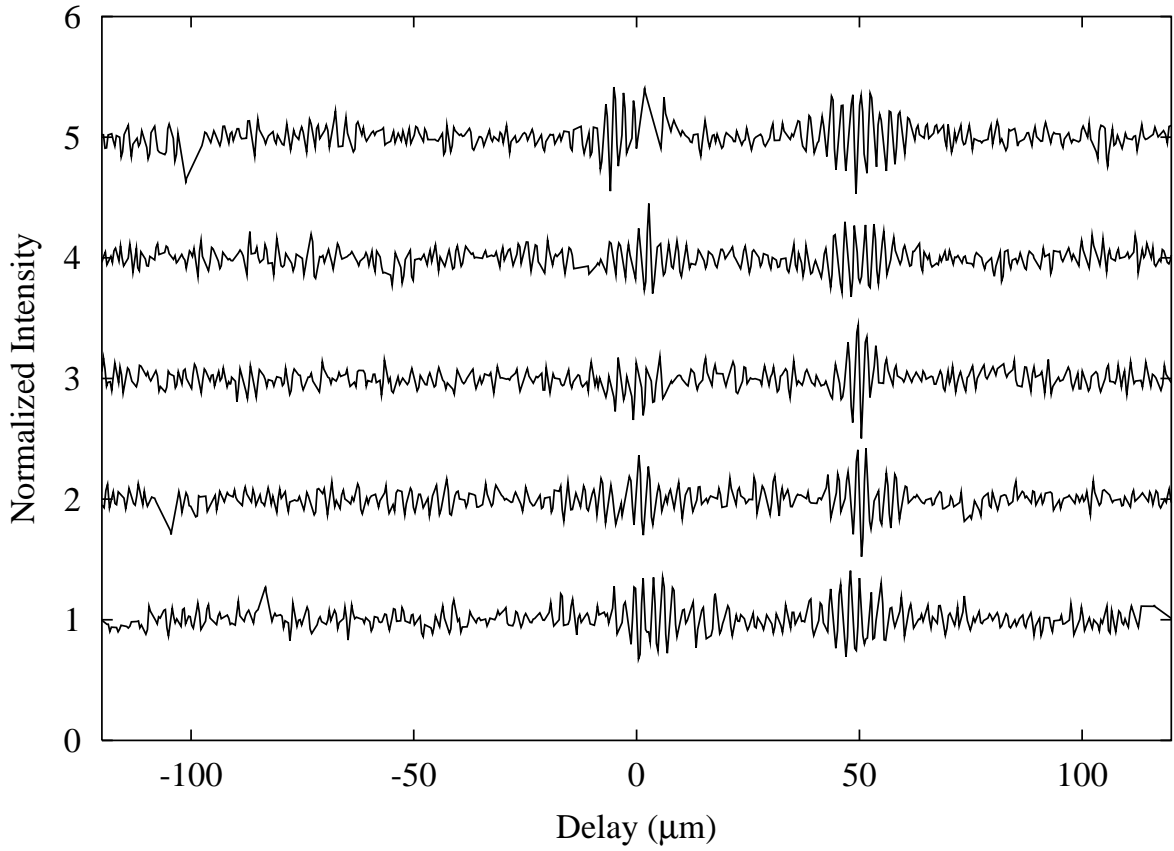}
\end{center}
\caption{\label{fig:fringe} A plot showing five consecutive filtered fringe scans of HD 171779.
Each scan corresponds to $\sim 1.5$ seconds of data. The scans have been normalized, 
and offset for clarity.}
\end{figure}

The intensity vs. delay position measurements produced by the
interferometer were processed into astrometric measurements as
follows: (1) Detector calibrations (gain, bias and background) were
applied to the intensity measurements.  (2) The residual phase errors
from the primary fringe tracker were converted to delay and applied to
the data. Note that while the intensity measurements were spaced
regularly in time, and the delay scanned linearly in time, the
variable amount of delay correction applied from the fringe tracker
resulted in the intensity measurements being unevenly spaced in
delay. This somewhat complicated the downstream processing, in that
FFT-based algorithms could not be used. (3) The data were broken up
into ``scans'' either when the delay sweep changed direction or when
the fringe tracker lost lock. (4) For each scan, a power spectrum was
calculated using a Lomb-Scargle \citep{scargle82,press92} algorithm.
This spectrum provided an SNR estimate based on the ratio of the power
in and out of the instrument bandpass. Only the scans with an SNR
greater than unity were kept. (5) The intensity measurements were
optionally bandpass-filtered to remove the effects of atmospheric
turbulence changing the amount of light being coupled into the
detector. We note however that the final results of the fit (see
below) did not depend on whether or not the filter was applied; the
results differed by $<<1\sigma$ (the results shown were based on
unfiltered data). See Figure \ref{fig:fringe} for an example of
filtered fringes. (6) The differential delay of each scan was found by
evaluating the least-squares goodness-of-fit ($\chi^2$) parameter for
a fit of a double fringe packet to each scan, for the range of
plausible delay separations and mean delay values, and selecting the
separation corresponding to the $\chi^2$-minimum. See the discussion
below for how plausible ranges were determined. (7) Given a set of
delay separations as a function of time, a simple astrometric model
was fit to the data. The free parameters were separation and position
angle, or equivalently, differential right ascension and declination
(Figure \ref{fig:fit}).

In order to characterize any systematic dependence on assumptions made
in the data reduction, such as the shape of the fringe envelope, the effect of
dispersion, the effect of filtering and the level of the fringe
SNR-cutoff, we exhaustively varied these parameters and and re-fit the
data each time.  In all cases the effect was small compared to the
claimed uncertainties. This suggests that our results are robust
against systematic errors in the data processing algorithm; many
potential systematic error sources are eliminated due to the
differential nature of the measurement.

\subsection{Resolving Fringe Ambiguity}

The oscillatory nature of the fringe means that there will be many
local minima in the $\chi^2$ surface, separated from the global
minium (at the center of the fringe packet) by integer multiples of
the wavelength $\lambda$.  In the presence of noise it is not always
obvious which is the global minimum.  In previous work
\citep{dyck95} this ambiguity was avoided by forming an ``envelope''
of the fringe packets, then fitting these envelopes. However, this
effectively removed the high-resolution phase information, which is
what provides the high precision astrometry.

We find that it is possible to co-add the $\chi^2(\Delta d)$ functions
from many scans. This increases the signal-to-noise ratio and so
reduces the likelihood of picking the ``wrong'' fringe.  In order to
perform the co-adding, the two-dimensional (separation and mean delay)
$\chi^2$ surface was projected into a one-dimensional space of delay
separation by selecting the best (lowest $\chi^2$) mean delay position
for each possible delay separation. The minimum of this co-added
$\chi^2$ function was used to determine the fringe separation to
within $\lambda$; the range of plausible separations (used in step 6
above) was limited to $\pm \lambda/2$ around this value.

The number of scans that could be co-added in this manner depended on
the rate of change of delay separation, with too long of an
integration time ``smearing'' the fringes. We found that in the case
of our observations, co-adding all the $\chi^2$ functions from a
75-second period gave adequate SNR to reliably (85--95\% of the time)
determine the central fringe. The remaining 5--15\% of scans produced
delays that were shifted off the median fringe position by a
wavelength. However, rather than artificially adjusting the delays by
a multiple of the wavelength, these scans were simply discarded. Note
that no {\em a priori} information was used to constrain the location
of the central fringe, and each 75-second group was treated
independently.

 We also point out that it should be possible to co-add $\chi^2$
functions not as a function of separation but as a function of
differential right ascension and declination; such an approach should
allow one to use the co-adding approach with very long integration
times (e.g. for all the scans of a given star in a night). We are
currently developing such an approach.

\section{Results}

\begin{figure}[]
\begin{center}
\plotone{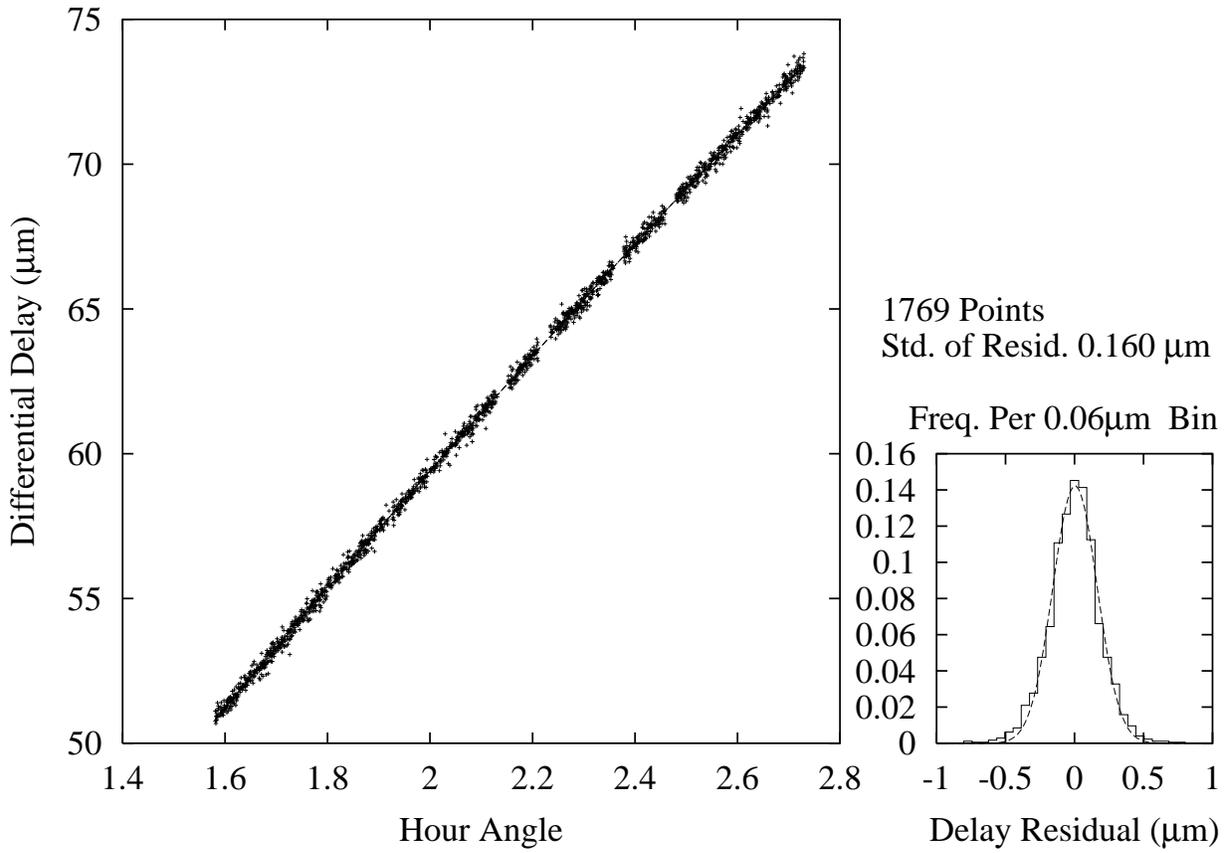}
\end{center}
\caption{\label{fig:fit} A fit of an astrometric separation model to the measured separations
as a function of time. Also shown is a histogram of the fit residuals,
which are well modeled by a Gaussian distribution.}
\end{figure}

\begin{figure}[]
\begin{center}
\includegraphics[scale=1.0]{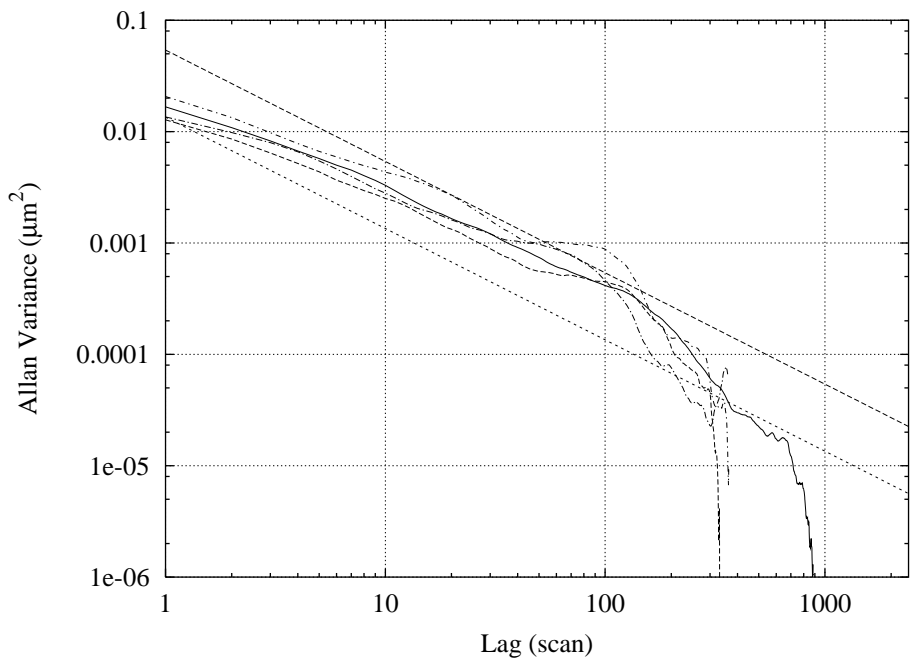}
\end{center}
\caption{\label{fig:allan} Allan variances of the residuals to the
astrometric fits, as well as a line showing the expected behavior
corresponding to white noise at a level equivalent to a fit precision
of $10~\mu$as (upper) and $5~\mu$as (lower) in in 1 hour of
integration (1.5 second scans).}
\end{figure}

\begin{figure}[]
\begin{center}
\plotone{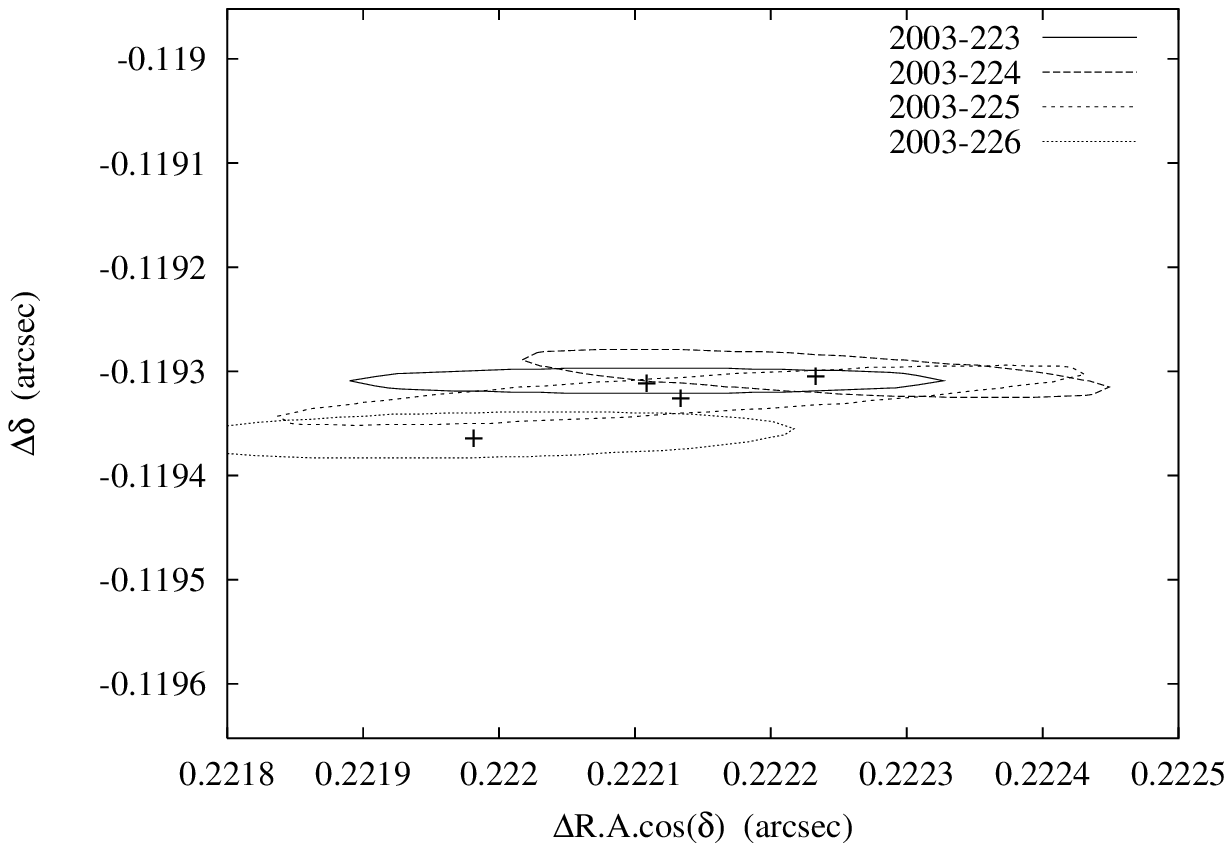}
\end{center}
\caption{\label{fig:error}A contour plot of the $\chi^2$ surfaces as a
function of the fit parameters, for data from four nights. The
1-$\sigma$ contours and the preferred solutions are marked.}
\end{figure}

We show the results of an astrometric fit to 45 minutes of data (taken
over the course of 70 minutes of observation) in Figures \ref{fig:fit}
and \ref{fig:error}. With 1769 scans used, we find the residual delay
errors to be well modeled by a Gaussian distribution with a full-width
at half-maximum of 0.160 $\mu$m.  In order to characterize the
residuals, and in particular determine if they could be considered to
be independent, we plot the Allan variance of the residuals in Figure
\ref{fig:allan}. The Allan variance \citep{tms01} at lag $l$ is given
by
\begin{equation}
\sigma^2_A(l) = \frac{1}{2(M' + 1 - 2l)}\sum_{n=0}^{M'-2l}\left ( \frac{1}{l} \sum_{m=0}^{l-1} x_{n+m} - x_{n+l+m}  \right )^2 
\end{equation}
\noindent where $M'$ is the total number of data points. As can be
seen in the figure, the residuals are white out to lags of more than
500 scans, implying a final astrometric precision of
10~$\mu$as. We list the results from 4 nights of observation
in Table \ref{tab:res}.

\begin{table}
 
\begin{tabular}{lcccc}
\tableline
\tableline
Epoch of Observation  & $\Delta$ R.A. & $\Delta \delta$ & \# Scans & $rms$\\
(MJD)                 & (arcsec)      & (arcsec)        &         &  $\mu$m \\
\tableline
52862.30085 & $ 0.222109 \pm 1.43\times10^{-4} $ & $ -0.119309 \pm 8.39\times10^{-6}$ & 1769 & 0.160\\
52863.28229 & $ 0.222233 \pm 1.43\times10^{-4} $ & $ -0.119302 \pm 1.55\times10^{-5}$ & 668 & 0.162\\ 
52864.31096 & $ 0.222134 \pm 1.97\times10^{-4}$ & $ -0.119323 \pm 1.89\times10^{-5}$ & 724 & 0.163\\
52865.29921 & $ 0.221981 \pm 1.55\times10^{-4}$ & $ -0.119361 \pm 1.48\times10^{-5}$ & 731 & 0.200 \\ 
\tableline
\end{tabular}
\caption{\label{tab:res} The results from four nights of differential astrometry 
of the binary HD 171779.}

\end{table}

The 1-sigma error region (found by plotting the $\chi^2 = \chi^2_{min}
+ \Delta\chi^2$ contour, Press et al. 1992\nocite{press92}) is highly
elliptical with the major axis oriented roughly parallel to the
R.A. axis. Such error ellipses are to be expected in single-baseline
interferometric data, which has limited sensitivity in the direction
perpendicular to the baseline. It should be noted however that for
sufficiently long observations, Earth-rotation will provide an
orthogonal baseline. The major and minor axes of the uncertainty
ellipse are easily found by diagonalizing the covariance matrix: the
magnitude of the uncertainty in the direction of the minor axis was
8.3 $\mu$as for the 10 August data and 12~$\mu$as for
the 11 August data; consistent with the delay residuals. The
uncertainty in the major axis direction was 144 and 143
~$\mu$as respectively. Figure \ref{fig:error} shows the four
error ellipses superimposed. We fit the measured differential
declinations to a linear trend of $-18 \pm 7~\mu$as/day; the
$r.m.s.$ of the residuals is $16~\mu$as, and the reduced
$\chi^2 = 1.1$.  The predicted change in differential declination due
to orbital motion is approximately $-11~\mu$as/day.

\section{Discussion}

We have achieved an astrometric measurement precision of $\approx 9$
$\mu$as in 70 minutes of observation on the 0.25
arcsecond separation binary star HD 171779. The measurements were
repeated on four nights, yielding answers consistent at the level of
$16~\mu$as. These early results show the promise of very
narrow-angle astrometry for achieving very high precision measurements
of stellar binaries with a separations on the order of 0.05--1
arcseconds.  This is particularly useful in searching for planets
around such binaries, although the resulting high-precision visual
orbits for these stars should also be useful in providing stellar
masses and distances at the level of $10^{-3}$ . It should be noted
that such high precision mass determinations may require an extended
period of observation as these systems can have very long orbital
periods.  It is also necessary to obtain radial velocity data accurate
to 10 m/s or better, which can be challenging.

This newly demonstrated approach to narrow-angle astrometry compares
quite favorably with other astrometric methods used to date, such as
speckle interferometry \citep{hor02,woi03}, single-telescope
astrometry with \citep{l01} and without \citep{ps96,ps00}
adaptive-optics, and space-based astrometry using HST
\citep{ben02}. These have all produced measurement precisions on the
order of 0.1-1 mas. The improved performance of the interferometric
approach is primarily due to the longer available baseline. However,
the combination of phase-referencing and fringe-scanning also compares
favorably to other interferometric methods, such as using fringe
visibility measurements to find binary star orbits ( i.e. a fractional
precision of $10^{-4}$ as compared to $10^{-2}$, Boden et al.,
2000\nocite{boden00}).  In this case the improvement comes from the
unbiased nature of the phase estimator.

Recent years have seen a veritable explosion in the number of known
extra-solar planets \citep{Sch:03::}, starting with the famous example
of 51 Peg \citep{May:95}. To date, most of these systems have been
found using high-precision radial velocity techniques, although recent
photometric transit searches have observed several planets
\citep{char00,kon03}. However, these methods have to a large extent
avoided searching close (separation less than a few arcseconds) binary stellar
systems, primarily because the techniques used are not well suited to
such systems. Therefore it is particularly interesting to note that
despite the deliberate avoidance of binaries, of the more than 100
known planets, 14 of them are in wide binary stellar systems. In
addition, given the high frequency of binary stellar systems (57\%
among systems older than 1 Gyr, \cite{Duq:91}), it is clear that any
comprehensive planetary census must address the question of how
frequently planets occur in such systems. This is all the more
relevant given that several theoretical investigations have indicated
that there exist regions in binary parameter space where planets can
form and exist in stable orbits over long periods of time, though this does
remain controversial \citep{Whi:98::,Bos:98::,Mar:00::,Nel:00::,Bar:02::}.

An instrument such as PTI, capable of 10 $\mu$as
very-narrow-angle astrometry, could be used to search many of the
brightest speckle binary systems for planets. We have compiled a list
of approximately 50 suitable systems, with $m_K < 4.5$, 
separations less than 1 arcsecond, and within the field of regard of
PTI. The median orbital separation between the binary components in
these systems is 19 AU, and hence there should be regions where
planets can remain stable for long periods. In particular, adopting
the result from \cite{holman99} we calculate the largest stable orbit
in each system.  We find that the median detectable planetary mass in
such an orbit is 0.5 Jupiter masses (assuming $3 \sigma$ confidence
detections).  The corresponding median orbital period is 2.2 years. A
limited survey could quickly begin to provide useful constraints on
the frequency of planets in binary stellar systems.  As a new
generation of long-baseline optical interferometers become operational
in the next few years, this type of survey could be easily extended to
sample sizes of several hundred stars, hence providing strong
constraints on planetary formation in the binary environment.

\acknowledgements We would like to thank E. Bertschinger, A. Boden,
B. F. Burke, M. Colavita, M. Konacki, S. R. Kulkarni \& N. Safizadeh
for their contributions to this effort. We also particularly
acknowledge the extraordinary efforts of K. Rykoski, whose work in
operating and maintaining PTI is invaluable and goes far beyond the
call of duty.  Observations with PTI are made possible through the
efforts of the PTI Collaboration, which we acknowledge. Part of the
work described in this paper was performed at the Jet Propulsion
Laboratory under contract with the National Aeronautics and Space
Administration. Interferometer data was obtained at the Palomar
Observatory using the NASA Palomar Testbed Interferometer, supported
by NASA contracts to the Jet Propulsion Laboratory. This research has
made use of the Simbad database, operated at CDS, Strasbourg,
France. MWM acknowledges the support of the Michelson Graduate
Fellowship program. BFL acknowledges support from a Pappalardo
Fellowship in Physics.

\end{document}